\documentclass{llncs}

  \title{Proof Hints for Event-B}
\author{Thai Son Hoang}
\institute{Institute of Information Security, ETH Zurich}

\usepackage[T1]{fontenc}
\usepackage[ansinew]{inputenc} 

\usepackage{abbrev}

\usepackage{url}

\usepackage[colour,small]{eventB}

\usepackage{bsymb}

\usepackage{rodin}

\usepackage{xspace}

\usepackage{space}

\usepackage{form-model}

\usepackage{tikz}

\usepackage{amsmath}

\usepackage{seqcal}

\usepackage[plainpages=false]{hyperref}

\hypersetup{
  pdftitle={Proof Hints for Event-B},
  pdfauthor={
    Thai Son Hoang
    Institute of Information Security, ETH Zurich, Switzerland
  },
  pdfsubject={Event-B, Proof Hints},
  pdfkeywords={Event-B, Proof Hints, Tool},
  colorlinks=true
}

\newcommand{\PREAMBLE}{}
\begin{document}
\maketitle

\begin{abstract}
  Interactive proofs are often considered as costs of formal modelling
  activity.  In an incremental development environment such as the
  Rodin platform for Event-B, information from proof attempts is
  important input for adapting the model.  This paper considers the
  idea of using interactive proofs to ``improve'' the model, in
  particular, to convert them into automatic ones.  We propose to lift
  some essential proof information from the interactive proofs into
  the model as what we called \emph{proof hints}.  In particular,
  proof hints are not only for the purpose of proofs: it helps to
  understand the formal models better.

\end{abstract}

%%% Local Variables: 
%%% mode: latex
%%% TeX-master: "proof-hints"
%%% End: 

\section{Introduction}
\label{sec:introduction}

\eventB is a refinement-based modelling method, which can be used to
develop various types of systems.  Starting with an abstract
specification, several refinement steps gradually introduce more
details into the formal models in a consistent manner.  An important
part of an \eventB formal model is the verification conditions
generated as proof obligations.  The task of discharging these
obligations is first given to some automated provers.  Afterwards,
remaining undischarged proof obligations are required to be proved
interactively.  Typically, manual proofs are considered as ``costs''
of development, given the required human interaction for produced
them, and the difficulty in maintaining them when the formal models
evolve.

As the size of models grows, the complexity of the associated proof
obligations also increases, hence interactive proofs are inevitable.
Improving the performance of the automatic provers has been considered
with some
success~\cite{roeder10:_relev_filter_event_b,schmalz11:_expor_isabel,DBLP:conf/asm/DeharbeFGV12}.
Despite the improvement in the percentage of automatic proofs,
interactive proofs are still an obstacle in developing and maintaining
formal models.

In this paper, we attempt to answer the following question. \emph{Can
  we improve our formal models in such a way that helps the proofs?}
After all, modelling using refinement is also a way of structuring the
proof of correctness of the models.  We propose some notions to
encapsulate essential proof details extracted from interactive proofs
within the formal model.  We call the additional information to the
formal models ``proof hints''.

Some form of \emph{proof hints} already exists in \eventB, \EG
``witnesses'' or ``theorems''.  These useful features are designed not
only to help with proving the correctness of the model but also to
give more information about the particular model, i.e., \emph{why it
  is correct}.  In fact, ``proof hints'' should help to understand the
formal model better.

We consider the current state of \Rodin and show two kinds of useful
proof information that can be included in the formal models, namely,
to \emph{select hypotheses} and to \emph{perform a proof by cases}.
\begin{description}
\item[Select hypotheses] Indicates that some facts (e.g., invariants or
  axioms) are required for discharging the obligations.
\item[Perform a proof by cases] Indicates that the proof can be discharged by
  consider different cases.
\end{description}
We show that the effect of the proof hints can be ``simulated'' at the
moment by some modelling ``tricks'' in \eventB.

In the long term, we propose to have an extension to \eventB and to
\Rodin, to have proof hints as a part of the model and to implement a
plug-in for interpreting the proof hints and applying these hints
appropriately during proofs.

% The question is how can we decide which ``features'' are useful to have in the
% model.  In principle we can go to the extreme and include the whole proof
% strategy into the model in order to be able to have it being discharged
% automatically.  But this is undesirable since it will make the model difficult
% to understand.  Having taken this into account we specify our criteria here for
% ``proof hints''.

% \begin{enumerate}
% \item It should help to understand the model better.

% \item It should improve the automatic proving rate of the model.
% \end{enumerate}

% In fact, we regard the first criterion as more important whereas the second
% criterion can be considered as a bonus for improving the clarity of the model.

\textit{Organisation}. The rest of the paper is structured as follows.
Section~\ref{sec:background} gives some background on \eventB and
\Rodin.  Section~\ref{sec:contributions} presents our ideas of proof
hints.  Section~\ref{sec:examples} illustrates proof hints by means of
two examples. Section~\ref{sec:tool-support} discusses some proposals
for the tool support.  We give some conclusions in
Section~\ref{sec:conclusion}.

%%% Local Variables: 
%%% mode: latex
%%% TeX-master: "proof-hints"
%%% End: 

\section{Background}
\label{sec:background}

\subsection{The Event-B Modelling Method}
\label{sec:event-b-modelling}

An \eventB model corresponds to a discrete transition system and is
divided into two parts: a \emph{static} part called \emph{context} and
a \emph{dynamic} part called \emph{machine}.  A context may contain
\emph{carrier sets} (types), \emph{constants}, \emph{axioms}
(assumptions about sets and constants).  For clarity, we omit
references to context in the sequel.

Machines may contain \emph{variables}, \emph{invariants}, and
\emph{events}. Variables $\absVrb$ define the state of a machine and
are constrained by invariants $\absInv(\absVrb)$.  An event
$\Bevt{\absEvt}$ can be represented as
% \begin{equation}
$\inlineevent{\absEvt}{\absPar}{\absGrd(\absPar,\absVrb)}{\absAct(\absPar,\absVrb)}$,
% \end{equation}
where $\absPar$ stands for the event's \emph{parameters}, thus
allowing for state changes. The \emph{guard}
$\absGrd(\absPar,\absVrb)$ states the necessary condition under which
an event may occur.  The \emph{action} $\absAct(\absPar,\absVrb)$
describes how state variables $\absVrb$ evolve when the event occurs.
The short form
% \begin{equation}
$\inlineevent{\absEvt}{}{\absGrd(\absVrb)}{\absAct(\absVrb)}\label{eq:event-when}$
% \end{equation}
is used when the event does not have any parameters, and we write
% \begin{equation}
$\inlineevent{\absEvt}{}{}{\absAct(\absVrb)}\label{eq:event-begin}$
% \end{equation}
when, in addition, the event's guard equals \emph{true}. A dedicated
event in the last form is used for the \emph{initialisation} event
($\INITIALISATION$).  The action of an event is composed of one or
more \emph{assignments} of the form: $\absVrb \bcmeq
\Expr(\absPar,\absVrb)$, $\absVrb \bcmin \Expr(\absPar,\absVrb)$,
$\absVrb \bcmsuch \Pred(\absPar,\absVrb,\absVrb^\prime)$, specifying
$\absVrb$ becomes $\Expr(\absPar,\absVrb)$, $\absVrb$ becomes an
element of $\Expr(\absPar,\absVrb)$, and $\absVrb$ becomes such that
$\Pred(\absPar,\absVrb,\absVrb^\prime)$ holds, respectively.  All
assignments of an action $\absAct(\absPar,\absVrb)$ occur
\emph{simultaneously}.  As a result, each event $\absEvt$ is
associated with a before-after predicate, denoted as
$\absBAP(\absPar,\absVrb,\absVrb^\prime)$.

The \emph{invariant preservation} proof obligation ($\Bpo{INV}$)
states that invariants are maintained whenever variables are updated.
For each event $\absEvt$, the corresponding proof obligation is as follows.
\begin{equation*}
  \footnotesize
  \sequent{
    \absInv(\absVrb),
    \absGrd(\absPar,\absVrb),
    \absBAP(\absPar,\absVrb,\absVrb^\prime)
  }
  {\absInv(\absVrb^\prime)}
  \label{po:inv}
  \tag{$\Bpo{INV}$}
\end{equation*}
  
All predicate modelling elements, \EG axioms, invariants, guards, can
be also declared as theorems.  Theorems need to be proved when they
are declared.  As an example, a theorem in guard must be proved to be
a consequence of axioms, invariants, and previously declared guards.

\emph{Machine refinement} is a mechanism for introducing details about
the dynamic properties of a model \cite{abrial10:_model_in_event_b}.
The states of the abstract machine $\absMch$ are related to the states
of the concrete machine $\cncMch$ by \emph{gluing invariants}
$\cncInv(\absVrb,\cncVrb)$, where $\absVrb$ and $\cncVrb$ are the
variables of the abstract and concrete machine, respectively.  Each
event $\absEvt$ of the abstract machine is \emph{refined} by a
concrete event $\cncEvt$ (later we will relax this constraint).
Assume that the concrete event is of the following form
$\inlineevent{\cncEvt}{\cncPar}{\cncGrd(\cncPar,\cncVrb)}{\cncAct(\cncPar,\cncVrb)}$.
Somewhat simplifying, we can say that $\absEvt$ refines $\cncEvt$ if
the gluing invariant establish a simulation of $\cncEvt$ by the
$\absEvt$.  This is presented as the following obligation.
\begin{equation*}
  \footnotesize
  \sequent{
    \absInv(\absVrb),
    \cncInv(\absVrb,\cncVrb),
    \cncGrd(\cncPar,\cncVrb),
    \cncBAP(\cncPar,\cncVrb,\cncVrb^\prime)
  }
  {\exists\absPar,\absVrb^\prime\qdot\absGrd(\absPar,\absVrb) \land
    \absBAP(\absPar,\absVrb,\absVrb^\prime) \land
    \cncInv(\absVrb^\prime,\cncVrb^\prime)}
  \label{po:ref}
  \tag{$\Bpo{REF}$}
\end{equation*}

In order to split the above proof obligation, \eventB introduces the
notion of ``witnesses'' for $\absPar$ and $\absVrb^\prime$.  The
witnesses are predicates
$\witness_1(\absPar,\cncPar,\absVrb,\cncVrb,\cncVrb^\prime)$ (for
$\absPar$) and
$\witness_2(\absVrb^\prime,\cncPar,\absVrb,\cncVrb,\cncVrb^\prime)$
(for $\cncVrb^\prime$), which are required to be \emph{feasible},
i.e., satisfying the following proof obligations.
\begin{equation*}
  \sequent{
      \absInv(\absVrb),
      \cncInv(\absVrb,\cncVrb),
      \cncGrd(\cncPar,\cncVrb)
    }{
      \exists \absPar \qdot \witness_1(\absPar,\cncPar,\absVrb,\cncVrb,\cncVrb^\prime)
    }
    \label{po:wfis1}
  \tag{$\Bpo{WFIS1}$}
\end{equation*}
\begin{equation*}
  \sequent{
      \absInv(\absVrb),
      \cncInv(\absVrb,\cncVrb),
      \cncGrd(\cncPar,\cncVrb),
       \cncBAP(\cncPar,\cncVrb,\cncVrb^\prime)
    }{
      \exists \absVrb^\prime \qdot \witness_2(\absVrb^\prime,\cncPar,\absVrb,\cncVrb,\cncVrb^\prime)
    }
    \label{po:wfis2}
  \tag{$\Bpo{WFIS2}$}
\end{equation*}
The witnesses supply instances of $\absPar$ and $\absVrb^\prime$
(provided that they exist) for instantiating the proof obligation
\ref{po:ref}.  Given the witnesses, the refinement proof obligation
\ref{po:ref} can be split into the following proof obligations.
\begin{equation*}
  \footnotesize
  \sequent{
      \absInv(\absVrb),
      \cncInv(\absVrb,\cncVrb),
      \cncGrd(\cncPar,\cncVrb),
      \witness_1(\absPar,\cncPar,\absVrb,\cncVrb,\cncVrb^\prime)
    }{\absGrd(\absPar,\absVrb)}
    \label{po:grd}
    \tag{$\Bpo{GRD}$}
  \end{equation*}
  \begin{equation*}
  \footnotesize
    \sequent{
      \absInv(\absVrb),
      \cncInv(\absVrb,\cncVrb),
      \cncGrd(\cncPar,\cncVrb),
      \cncBAP(\cncPar,\cncVrb,\cncVrb^\prime),
      \witness_1(\absPar,\cncPar,\absVrb,\cncVrb,\cncVrb^\prime),
      \witness_2(\absVrb^\prime,\cncPar,\absVrb,\cncVrb,\cncVrb^\prime)
    }{\absBAP(\absPar,\absVrb,\absVrb^\prime)}
    \label{po:sim}
    \tag{$\Bpo{SIM}$}
  \end{equation*}
  \begin{equation*}
  \footnotesize
    \sequent{
      \absInv(\absVrb),
      \cncInv(\absVrb,\cncVrb),
      \cncGrd(\cncPar,\cncVrb),
      \cncBAP(\cncPar,\cncVrb,\cncVrb^\prime),
      \witness_1(\absPar,\cncPar,\absVrb,\cncVrb,\cncVrb^\prime),
      \witness_2(\absVrb^\prime,\cncPar,\absVrb,\cncVrb,\cncVrb^\prime)
    }{\cncInv(\absVrb^\prime,\cncVrb^\prime)}
    \label{po:inv-ref}
    \tag{$\Bpo{INV}$}
  \end{equation*}

% New event
In the course of refinement, \emph{new events} are often introduced
into a model. New events must not modify abstract variable $\absVrb$.
% Split 
When an abstract event $\absEvt$ is refined by more than one
concrete events $\cncEvt$, we say that the abstract event
$\absEvt$ is \emph{split} and prove that each concrete $\cncEvt$
is a valid refinement of the abstract event.
% Merge
A concrete event $\cncEvt$ can refine two (or more) abstract events
$\absEvt$, provided that the abstract events are only different in
guards.  We say that the abstract events are \emph{merged} into the
concrete event $\cncEvt$.  It is required to prove that the guard of
$\cncEvt$ is stronger than the disjunction of the guards of the
abstract events.

\subsection{Proving with the Rodin Platform}

\Rodin is an Eclipse-based tool chain for analysing and reasoning
about \eventB models.  Models are developed incrementally within the
platform.  Two main activities of developers are \emph{modelling} and
\emph{proving} (as illustrated in \Fig{fig:developing}).  Proof
obligations are generated from modelling and are input to the proving
activity.  Information about proof attempts from proving are input to
the modelling activity to ``improve'' the models.  In particular,
failed proof attempts usually indicate some problems with the models
and give hints on how the models can be fixed, \EG to strengthen the
guard of some events or to add some missing invariants into the
models.
\begin{figure}[!htbp]
  \footnotesize
  \centering
  \ifx\PREAMBLE\UnDef
\documentclass{beamer}
\usepackage{tikz}
\usepackage[english]{babel}
% or whatever

\usepackage[latin1]{inputenc}
% or whatever

\usepackage[T1]{fontenc}
\usepackage{amssymb}
\usepackage{amsmath}
\usepackage{eventB}

\begin{document}
\else
\fi

\begin{tikzpicture}[scale=0.6]

  \draw (0, 0) node[draw, minimum width=6em, minimum height=3em](modelling){Modelling};
  \draw (10, 0) node[draw, minimum width=6em, minimum
  height=3em](proving){Proving};

  \draw[thick,->] (modelling) .. controls (3,2) and (7,2)
  .. node[midway, sloped, above]{proof obligations} (proving) ;

  \draw[thick,->] (proving) .. controls (7,-2) and (3,-2)
  .. node[midway, sloped, below]{proof attempts} (modelling) ;
\end{tikzpicture}

\ifx\PREAMBLE\UnDef
\end{document}
\else
\fi
  \caption{Developing \eventB models using \Rodin}
  \label{fig:developing}
\end{figure}

% Automatic v.s. Interactive proofs.
Obligations are proved either automatically or manually.  In automatic
mode, \Rodin uses some predefined proof tactics made up of internal
and external provers to discharge the obligations.  In interactive
mode, the user ``guides'' the proof attempts by applying some simple
proof steps to simplify the obligations before invoking some trusted
external provers to finish the proofs.  As interactive proofs require
manually intervenions, it is usually considered as some costs of
developing formal models.  Moreover, maintenance of interactive proofs
is difficult: a change in the formal model more often invalidates the
interactive proofs.  As a result, improving the rate of automatic
proofs will also help to maintain the models better.

% Some common interactive proof steps.
We consider some common interactive proof steps, e.g., to \emph{add
  hypothesis}, to \emph{select hypotheses}, and to \emph{perform a
  proof by cases}.
\begin{description}
\item[Add hypothesis] This proof step corresponds to the following
  proof rule. 
  \begin{Bcode}
    $
    \proofrule{CUT}{
      \sequent{\Hbold}{P} ~~~\sequent{\Hbold, P}{G}
    }{\sequent{\Hbold}{G}}
    $
  \end{Bcode}
  The rule allows add $P$ as a hypothesis, provided that it can be
  proved from the current hypotheses $\Hbold$.

\item[Select hypotheses] \Rodin has a notion of \emph{selected}
  hypotheses which is used by some external provers.  Often, too many
  irrelevant hypotheses will have negative effect on the performance
  of external provers.  By restricting the set of hypotheses available
  to these external provers, the user helps the external provers to
  concentrate on using only some relevant hypotheses.  An example for
  selected hypotheses are guards of an event: they are by default
  selected for proof obligations related to the event.

\item[Perform a proof by cases] This proof step allows user to perform
  a proof by cases, with respect to some condition $P$.
  \begin{Bcode}
    $
    \proofrule{CASE}{
      \sequent{\Hbold, P}{G} ~~~\sequent{\Hbold, \neg P}{G}
    }{\sequent{\Hbold}{G}}
    $    
  \end{Bcode}
  The proof is split into two branches accordingly.
\end{description}

%%% Local Variables: 
%%% mode: latex
%%% TeX-master: "proof-hints"
%%% End: 

\section{Proof Hints}
\label{sec:contributions}
% Existing work on external provers
There are existing work for improving the rate of automatic proofs.
Recently, some links to external provers such as
Isabelle~\cite{schmalz11:_expor_isabel},
SMT~\cite{DBLP:conf/asm/DeharbeFGV12} have been added to \Rodin.  Selected
hypotheses can be calculated according to some
heuristic~\cite{roeder10:_relev_filter_event_b}.  However, interactive
proofs are still unavoidable.  We look at the problem from a different
angle: to convert interactive proofs into automatic proofs by
improving the formal models, essentially exposing some proof
information in the formal models.  We call these additional proof
information \emph{proof hints}.

There is already several such proof information existing in the
\eventB models, normally being seen as part of the model rather than
some exposed proof information.
\begin{itemize}
\item Theorems in the model is a special case of \emph{adding a
    hypothesis} in an interactive proof.

\item Automatic \emph{selection of guards} for the event's proof obligations.

\item The witnesses can be seen as some hints for manually
  \emph{instantiating} the existential goal of the general proof
  obligation \ref{po:ref}, which results in three sub-obligations
  \ref{po:grd}, \ref{po:sim}, and \ref{po:inv}.
\end{itemize}

In principle, any proof information can be lifted to be proof hints,
part of the model.  However, revealing the actual proof is certainly
undesirable: this could have negative effects on the understanding of
the model.  In fact we want only to exposing \emph{essential}
information about the proofs.  We believe that the criteria for proof
hints should be as follows.
\begin{enumerate}
\item They should help to \emph{automate} more proofs.

\item They should help to better \emph{understand} the model.
\end{enumerate}

While the first criteria is straightforward, our emphasis is on the
second criteria.  Once again, low level proof information is
irrelevant for understanding of the model. We only want to have
essential key important proof steps as hints, in order to justify
about the correctness of the formal models.

%%% Local Variables: 
%%% mode: latex
%%% TeX-master: "proof-hints"
%%% End: 

\section{Some Useful Proof Hints}
\label{sec:examples}

This section presents two kinds of proof hints, namely to \emph{select
  hypotheses} and to \emph{perform a proof by cases}.%   For uniformity,
% each proof hint is presented according to the following structure.
% \begin{enumerate}
% \item Firstly, a brief description of the proof hints with the
%   situation where this can be helpful.

% \item Secondly, an example in \eventB is presented.  These example are
%   developed using version 2.4.0 of \Rodin with default setting for the
%   automatic provers.

% \item Thirdly, we show a solution (rather a workaround) for the
%   current \RODIN.

% \item Finally, we propose some extensions for the \eventB related to
%   the hint.
% \end{enumerate}

\subsection{Select Hypotheses}
\label{sec:hypotheses-selection}

During an interactive proof section, the developer can complete the
proof by selecting some hypotheses and invoke one of the provers that
uses only selected hypothesis, \EG AterlierB P0.  The solution to make
the proof become automatic is to (somehow) give hints to \Rodin to
select these additional hypotheses automatically.

\paragraph{Example}
Consider the following specification containing two variables $\XX$
and $\YY$.  The machine has a single event%
\footnote{%
  For clarity we omit the initialisation event $\INITIALISATION$.%
} called $\set$, which assign $\YY + 1$ to $\XX$ when $\XX$ is either
$1$ or $2$.
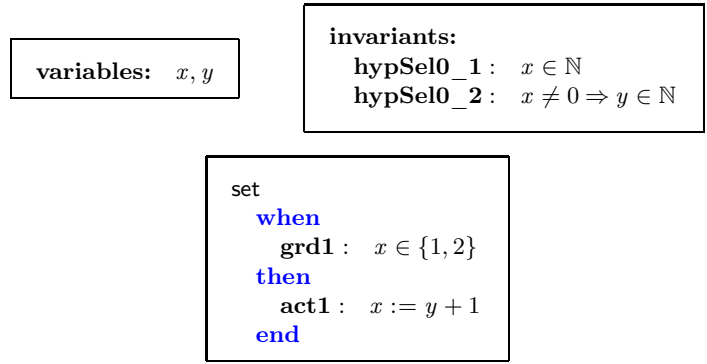
\begin{Bcode}
  $
  \variables{\XX, \YY}
  $
  \hspace{2em}
  $
  \invariants{
    \Binv{hypSel0\_1}: & \XX \in \nat \\
    \Binv{hypSel0\_2}: & \XX \neq 0 \limp \YY \in \nat \\[2ex]
  }
  $
  \\[2ex]
  $
  \event{\set}{}{\Bact{grd1}: & \XX \in \{1, 2\}}{\Bact{act1}: & \XX \bcmeq \YY + 1}
  $
\end{Bcode}
We are interested in proof obligation $\invpo{\set}{hypSel0\_1}$ stating
that event $\set$ maintains the invariant $\Binv{hypSel0\_1}$.
\begin{Bcode}
  $
  \sequent{
    \XX \in \nat,
    \XX \neq 0 \limp \YY \in \nat,
    \XX \in \{1, 2\}
  }{
    \YY + 1 \in \nat
  }
  $
\end{Bcode}
The proof obligation cannot be discharged automatically.  In
particular, by default, the selected hypotheses are
$\Binv{hypSel0\_1}$ and $\Bact{grd1}$.  The obligation can be
discharged by selecting $\Binv{hypSel0\_2}$, and invoke external
provers using selected hypotheses, such as AterlierB P0.

\paragraph{Workaround}
A simple workaround to have $\Binv{hypSel0\_2}$ being selected
automatically is to add the invariant as a \emph{theorem} in guard of
$\set$.
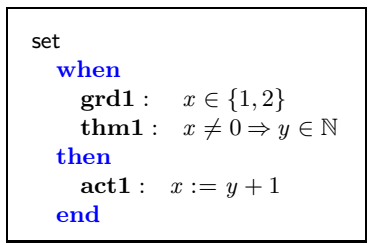
\begin{Bcode}
  $
  \event{\set}{}{
    \Bact{grd1}: & \XX \in \{1, 2\} \\
    \Bthm{thm1}: & \XX \neq 0 \limp \YY \in \nat    
  }{
    \Bact{act1}: & \XX \bcmeq \YY + 1
  }
  $
\end{Bcode}
The additional theorem can be removed in further subsequent refinement
if necessary.

\paragraph{Proposal}

The disadvantages of the above approach are as follows.
\begin{itemize}
\item This introduces extra proof obligations to prove that the copies
  of the invariants are theorems in guard (even though those proof
  obligations are discharged automatically).

\item Recopying the text of the invariants is error-prone.

\item Reformulating invariants leads to the need for changing the text
  of the extra theorems.
\end{itemize}

Our proposal is to have a specific ``proof hint'' for events.  This
form of proof hints will specify the invariant/theorem need to be used
automatically.  For example, we could extend event $\set$ with the
following hint of using $\Binv{hypSel0\_2}$ for the maintenance of
$\Binv{hypSel0\_1}$.
\begin{Bcode}
  $
  \begin{array}[!htbp]{|@{\quad}l@{\quad}|}
    \hline
    \set \\
    \quad \Bwhen\\
    \quad \quad \XX \in \{1, 2\} \\
    \quad \Bthen \\
    \quad \quad \XX \bcmeq \YY + 1  \\
    \quad \Bkeyword{hints}\\
    \quad \quad \mathsf{use} ~\Binv{hypSel0\_2}~\mathsf{for}~\Binv{hypSel0\_1} \\
    \quad \Bend \\
    \hline
  \end{array}
  $
\end{Bcode}

% This hints present the fact that while doing modelling, there exists some
% reasoning about the correctness of a particular events.  One can argue over
% having the hints specific for each proof obligation, but this does not effect
% much the performance of the automated provers or improve the clarity of the
% model.  Similar hints can be associated with theorems.

% For the example above, the event $\set$ can be as follows.  Here we assume
% that the model is name $\Bmch{m0}$.

\subsection{Perform a Proof by Cases}
\label{sec:proof-cases}

Sometimes, during an interactive proving session, the user suggests a
predicate $\Pred$ in order to do a ``proof by cases''.  The subsequent
branches of the proof can be discharged easily.  It is hence desirable
to have this hint about performing a proof by cases in the model.
Automatic provers often do not apply the case splits automatically,
since this potentially leads to blow up in terms of the number of
sub-goals.

\subsubsection{Example}
\label{sec:case-example}

Consider the following specification with three variables $\AA$,
$\BB$, $\CC$.
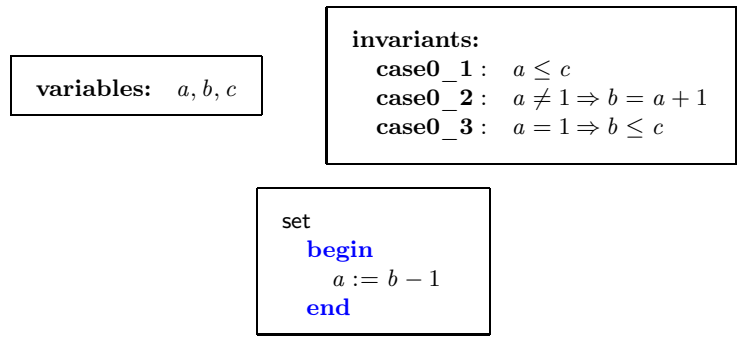
\begin{Bcode}
  $
  \variables{\AA, \BB, \CC}
  $
  \hspace{2em}
  $
  \invariants{
    \Binv{case0\_1}: & \AA \leq \CC \\ 
    \Binv{case0\_2}: & \AA \neq 1 \limp \BB = \AA + 1 \\ 
    \Binv{case0\_3}: & \AA = 1 \limp \BB \leq \CC \\[2ex] 
  }
  $
  \\[2ex]
  % $
  % \init{
  %   a \bcmeq 0 \\
  %   b \bcmeq 1 \\
  %   c \bcmeq 0
  % }
  % $
  % \hspace{2em}
  $
  \event{\set}{}{}{\AA \bcmeq \BB - 1}
  $
\end{Bcode}
The interesting proof obligation to look at is
$\invpo{\set}{case0\_1}$.  
\begin{Bcode}
  $
  \sequent{
    \AA \leq \CC,
    \AA \neq 1 \limp \BB = \AA + 1,
    \AA = 1 \limp \BB \leq \CC
  }{
    \BB - 1 \leq \CC
  }
  $
\end{Bcode}
Informally, the reasoning follows the cases of either $\AA = 1$ or
$\AA \neq 1$ and applying $\Binv{case0\_2}$, $\Binv{case0\_3}$
accordingly.  Hence we would like to give the hints about the case
split.  The obligation is not discharged by the default automatic
prover within \Rodin.

\subsubsection{Workaround}
\label{sec:case-workaround}
In order to ``simulate'' the effect of introducing this proof hints, we first
split the event into two sub-events, guarded by corresponding conditions.
\begin{Bcode}
  $
  \inlineevent{\set\_case1}{}{\AA = 1}{\AA \bcmeq \BB - 1}
  $
  \\[1ex]
  $
  \inlineevent{\set\_case2}{}{\AA \neq 1}{\AA \bcmeq \BB - 1}
  $
\end{Bcode}
The original event $\set$ can be obtained by merging the above two
events using refinement.
\begin{Bcode}
  $
  \revent{\set}{\set\_case1, \set\_case2}{}{}{}{\AA \bcmeq \BB - 1}
  $
\end{Bcode}
which leads to a trivial proof obligation to prove about merging events.

\subsubsection{Proposal}
\label{sec:case-proposal}

There are several disadvantages of the workaround:
\begin{itemize}
\item Splitting of event and merging using refinement is artificial.

\item Splitting of events leads to double number of proof obligations (those
  that does not need the case split).
\end{itemize}

Our proposal is to provide a hint directly in the model about the case split.
\begin{Bcode}
  $
  \begin{array}[!htbp]{|@{\quad}l@{\quad}|}
    \hline
    \Bevt{set} \\
    \quad \Bbegin \\
    \quad \quad \AA \bcmeq \BB - 1 \\
    \quad \Bkeyword{hints}\\
    \quad \quad \textsf{split case using} ~\AA = 1 ~\mathsf{for}~ \Binv{case0\_1} \\
    \quad \Bend \\
    \hline
  \end{array}
  $
\end{Bcode}

%%% Local Variables: 
%%% mode: latex
%%% TeX-master: "proof-hints"
%%% End: 

\section{Ideas on Tool Support}
\label{sec:tool-support}

Given the extensibility of \Rodin, having proof hints as additional
elements of \eventB models would be straightforward.  How the hints
are interpreted and work with the automatic provers of \Rodin is a
more challenging topic.  There are some options for the implementation
of the ``hint-interpreter''.

The first option is to have the interpreter to effect the generated
proof obligations.  For example, two different proof obligations are
generated according to the ``proof-by-cases'' hint to replace the
original proof obligation.  This requires to alter the \POG of \Rodin
to take into account the hints.  The second option is to leave the
original proof obligations untouched and to incorporate the hints at
the start of a proof, \IE they can be applied before the automatic
provers are invoked.

At the moment, we are investigating these options for tool support.
The first option is similar to witnesses in event refinement, to split
proof obligation \ref{po:ref} into obligations \ref{po:grd},
\ref{po:sim}, and \ref{po:inv}.  As a result, it changes the generated
proof obligations of \eventB models, which might be undesirable.  In
particular, the number of proof obligations generated can be different
depending on whether proof hints are present.  The second option
applying proof hints at the beginning of each proof, works for the two
illustrated proof hints presented within this paper.  In general, we
might want to have more general proof hints that should be apply in
the middle of a proof, or even combining different hints in one proof.

%%% Local Variables: 
%%% mode: latex
%%% TeX-master: "proof-hints"
%%% End: 

\section{Conclusion}
\label{sec:conclusion}
We presented some ideas about the notion of ``proof hints'' for
\eventB and discusses the possibilities of extending the supporting
\RODIN.  In a broad term, proof hints essentially are proof
information that are added to the model.  We proposed two kinds of
proof hints in this paper, for suggesting selected hypotheses and
performing proof-by-cases.  We presented some workaround at the moment
to ``simulate'' the proof hints and to automate some proofs in the
current version of \Rodin.  The illustrated examples are simple and
seem to be unrealistic.  However, they are extracted from some large
development (adapted accordingly).  Their simplicity is not too
illustrate the weakness of \Rodin's automatic provers, but rather to
support the argument that formal proofs of systems are challenging
tasks.

Often when describing an \eventB formal model, we also need to explain
why the model is correct, \EG why guard strengthening or invariant
preservation is satisfied.  Proof hints should give some information
(but not too much) to answer the questions about the correctness of
models.  It would be the ultimate goal of having self-explained formal
models, not only in terms of how they works (\EG events) and what
their properties are (\EG invariants), but also why they are correct.

We do not propose proof hints as a way to avoid interactive
proofs.  More often, we need to perform some interactive proof steps,
in order to figure out or understand why the obligation can be
discharged.  The role of proof hints therefore is to convert some
interactive proofs into automatic ones, helping the model to be more
resilient to changes.

In the long term, we might want to extract the essence of proofs as
some high-level structured proofs (similar to
Isabelle/Isar~\cite{wenzel02:_isabel_isar_refer_manual}).  This
requires more investigation in terms of the usefulness of such
an approach, and subsequent tool support.
 
%%% Local Variables: 
%%% mode: latex
%%% TeX-master: "proof-hints"
%%% End: 

\bibliography{proof-hints}

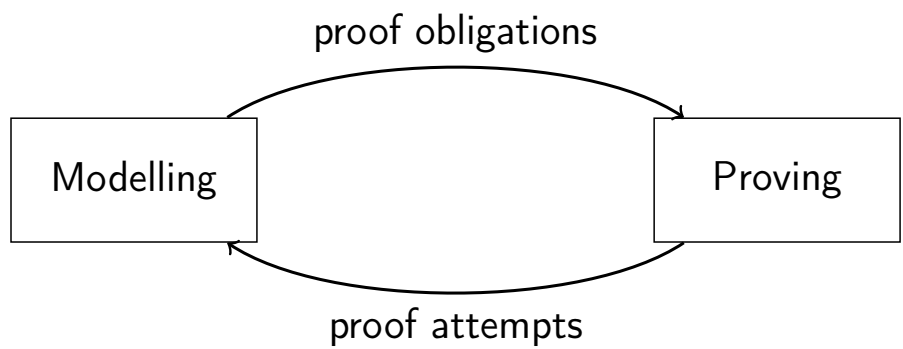
\end{document}